\documentclass[sigconf,authorversion,nonacm,linkcolor=blue,colorlinks]{acmart}

\usepackage[utf8]{inputenc} %
\usepackage[T1]{fontenc}    %
\usepackage{nicefrac}       %

\usepackage[colorinlistoftodos]{todonotes}
\usepackage{tcolorbox}

\newenvironment{example}[1][]{        
    \vspace{0mm}\begin{tcolorbox}[
        coltitle=black,
        colframe=white,
        left=1mm,
        right=1mm,
        top=1mm,
        bottom=1mm,
        boxsep=2mm,
        boxrule=0mm,
        arc=1mm,
    ]}
{\end{tcolorbox}\vspace{0mm}}

\newcommand{\todoLater}[1]{ 
}

\def\tightlist{} %

\title{`Generative CI' through Collective Response Systems%
}

\author{Aviv Ovadya}
\orcid{0000-0002-8766-0137}
\email{aviv@aviv.me} 
\affiliation{Technology and Public Purpose Fellow, Belfer Center, Harvard Kennedy School \country{USA}}
\affiliation{Visiting Scholar, Leverhulme Centre for the Future of Intelligence, University of Cambridge \country{UK}}

\begin{abstract}
This paper frames a specific kind of generative collective intelligence (CI) facilitation system: the \textbf{\emph{collective response system}}---and the \emph{collective dialogues} that it makes possible. It defines their structure, processes, key properties, and key principles with the goal of creating a useful shared language. It is intended to motivate the potential benefits of such systems and act as a concise reference text.

Collective response systems enable a form of ‘generative voting’---where both the `votes' \emph{and the choices of what to vote on} are provided by the collective.
This allows diverse populations to \emph{express} their perspectives and to \emph{hear} those of others---in ways that overcome the traditional limitations of polling, town halls, voting, referenda, etc. 
This can enable \emph{non-confrontational exploration of divisive issues}, can help \emph{identify common ground}, and \emph{surfaces insights} from those closest to the issues; thus overcoming gridlock around conflict and governance challenges, \emph{increasing trust}, and \emph{developing mandates}.\footnote{In other words, they help address both `alignment' and `optimization' problems in an environment with many options, conflicting values, and diverse stakeholders.}

 Notable examples of existing collective response systems include Polis and Remesh. Polis has been used \href{https://www.wired.co.uk/article/taiwan-democracy-social-media}{by Taiwan} and around the world for policy-making at different levels of government \cite{smallPolis2021}. The United Nations \href{https://unsmil.unmissions.org/asrsg-williams-conducts-digital-dialogue-1000-libyans}{deployed} Remesh, an AI-supported collective response system to understand the challenges and needs of ordinary people across a war-torn country---and with thousands of direct participants and a substantial proportion of the population engaging on social media \cite{masoodalavi2022UsingArtificialIntelligence, RemeshUN22MakingPeopleVoices, 2021ASRSGWilliamsConductsRemesh}.

Existing collective response systems build on technical work across recommender systems, language models, and human-computer interaction among other disciplines.\footnote{As with any social technology, the non-technical aspects are in many ways more important and impactful---but the technical components define and constrain key components of the medium. As this paper is particularly meant to be a concise overview for more technical researchers and practitioners, we focus on aspects most relevant to that purpose; however, a more exhaustive interdisciplinary exploration would be invaluable and there are many more fields that such an exploration would draw from. Please reach out to the author if interested in being involved in such a project.} Continued progress toward the development and adoption of such systems could help revitalize democracies, reimagine corporate governance, transform conflict, and govern powerful AI systems---both as a complement to deeper deliberative democratic processes and as an option where deeper processes are not possible.
\end{abstract}

\begin{document}

\maketitle


\hypertarget{motivation}{%
\section{Motivation}\label{motivation}}

Most current approaches for group understanding and decision-making don't work well at scale; particularly given the warped incentives of the modern attention economy: 

\begin{itemize}
\item
  \textbf{Thoughtful deliberation doesn't scale}: If a `town hall' of a thousand people all try to talk and listen simultaneously, the result is just noise. Alternatively, if people speak one by one, that can take days. With a million people, this would take decades for a single issue, so most people impacted by a policy or conflict cannot meaningfully speak or listen.
\item
  \textbf{People feel voiceless and disrespected:} People often don't feel empowered to express their perspectives, and don't see their views and experiences reflected. This creates mistrust and causes people to tune out of governance and conflict resolution processes.
\item
  \textbf{Valuable insights are missed}: `Knowledge keys'---information, ideas, and insights that can help overcome entrenched conflicts---are often missed. These often originate in those closest to an issue, many of whom have minimal reach through standard channels.
\item
  \textbf{Divisiveness wins over common ground:} Stirring up conflict is rewarded far more than identifying and elevating common ground.
\end{itemize}

However, connectivity, machine learning, and democratic practice advances\footnote{Specifically, improvements in collaborative filtering and language modeling, in combination with effective product design and facilitation insights.} may allow us to overcome some of these challenges with `simultaneous communication at a scale.'

Collective response systems are meant to enable groups of arbitrary scale to make generative decisions (e.g.~decisions where the participants develop the option space).    They are designed to get as close as possible to one version of the `democratic ideal'---that in collective decisions:

\begin{enumerate}
\def\labelenumi{\arabic{enumi}.}
\tightlist
\item
  Everyone has the ability to respond with their perspective
\item
  Everyone's response is listened to and incorporated.
\item
  The response(s) that best represent the group is chosen.\footnote{This might be e.g.~the response with the most total ```votes'', with the most total votes across divides (e.g.~like Polis \cite{smallPolis2021} and related systems with bridging-based ranking \cite{ovadya2022bridging}), or something very different. It could also be adjusted to compensate for differences in properties between a participating group and a larger group they represent (analogous to what is done in polling). }
\end{enumerate}

Collective response systems are decision and deliberation-focused, and aim to avoid the perverse incentives of entertainment or politics-focused communication environments (e.g.~social media) where attention-seeking behaviors are rewarded. %
They enable a form of focused simultaneous communication at scale. They also enable collectives to be treated as single agents, which can then interact with other such collective agents. %

\subsection*{Contributions}
This framing paper defines the structure, processes, key properties, and key principles of collective response systems. It is intended to provide a shared language for research and practice around collective intelligence facilitation and scaled democratic decision-making, informed by advances in human-computer interaction and artificial intelligence.

\hypertarget{structure-and-process}{%
\section{Structure and Process}\label{structure-and-process}}
A \textbf{collective response system} is a collective intelligence facilitation system that satisfies the structure, processes, properties, and principles described below, in sections \ref{structure-and-process} and \ref{system-design-properties-and-principles}.

\hypertarget{input}{%
\subsection{\textbf{Input}: A \emph{group} and a \emph{prompt}}\label{input}}

There are two inputs to the system: the \emph{prompt}\footnote{While the prompts in collective response systems are intended primarily for humans, this frame is similar in some ways to that of generative language models like GPT-3 \cite{brown2020GPT3LanguageModelsAre}. However, there are a number of ways in which generative machine learning models might play a role in such systems.} which might be a question, something to be completed, etc. and a \emph{group} which is the set of participants.

\begin{example}
\hypertarget{examples}{%
Examples:\label{examples}}

\begin{itemize}
\tightlist
\item
  For the citizens of a city: \emph{The most important challenge in education facing our city is\ldots{}}
\item
  For a platform like Facebook: \emph{What kinds of behavior should not be allowed on this platform?}
\item
  For students about to graduate: \emph{Who should be our graduation speaker?}
\item
  For an organization: \emph{The core values of our organization should be:}
\item
  For the global population: \emph{The most important steps we should be taking to address climate change are\ldots{}}
\end{itemize}
    
\end{example}

\hypertarget{output}{%
\subsection{\textbf{Output}: A \emph{representative distillation} of the evaluations and optionally the \emph{raw} and/or \emph{derived data}.\label{output}}}

The core outputs of such a collective response system are an \emph{approximation} or \emph{aggregation} of the responses that \emph{meaningfully} and \emph{representatively} convey the `best' response(s). This can most generally be referred to as a \textbf{representative distillation}.

The definition of `best' is largely open to interpretation but \emph{is} constrained by the \hyperlink{key-principles}{key principles described below}. %
For example, the responses with the most approval might be considered the best, though other approaches may be used instead (e.g. the \emph{bridging-based ranking} \cite{ovadya2022bridging} implemented by Polis's group informed consensus \cite{smallPolis2021} or by Twitter's Community 
Notes \cite{wojcikBirdwatchCrowdWisdom}). Outputs such as e.g.~the ``best response'' may also be used in downstream systems, including as \hyperlink{collective-dialogue}{described below} in future collective response processes.%

In addition, derived data useful for reflecting back the responses and evaluations of the group may be output. For example, systems like Polis can show participants how issues and people fit on a map of perspectives.
Finally, raw results such as the responses and evaluations may be output for further processing and analysis. %

\hypertarget{Process}{%
\subsection{\textbf{Process}: Responding, Evaluating, Distilling}\label{process}}

A process facilitated by a collective response system is called a \textbf{collective response process}. Such processes are made up of three kinds of subprocess:\footnote{These subprocesses might not be done strictly in a ``waterfall'' manner, but partially interlaced asynchronously depending on the system.}

\begin{itemize}
\tightlist
\item
  \textbf{Responding subprocess}: Everyone in the group can (optionally) respond to the prompt. 
\item
  \textbf{Evaluating subprocess}: Everyone in the group evaluates some subset of the responses (potentially chosen by the system).
\item
  \textbf{Distilling subprocess}: The system approximates and/or aggregates the evaluations of each response to produce a useful output.
\end{itemize}

\begin{example}
    
Putting this all together, consider the case of five thousand people in a city participating in a collective dialogue on education policy.

The \textbf{collective response process} in this case might look like this:
\begin{enumerate}
\def\labelenumi{\arabic{enumi}.}
\tightlist
\item
  \textbf{\emph{Responding:}} Participants are asked: \emph{``What is the most important challenge in education facing our city?''}
  \\ They can respond with a short answer---a response. In systems like Polis \cite{smallPolis2021} and Remesh \cite{RemeshWebsite} these responses are usually short, e.g.~one to three sentences.
\item
  \textbf{\emph{Evaluating}}: Participants are assigned \emph{`voting tasks'}, for example, to evaluate if they agree or disagree with a response.
\item
  \textbf{\emph{Distilling:}} The system can then \emph{show} the participants which responses have the most \emph{approval} and also the most \emph{common ground} across (e.g.~political) divides.
\end{enumerate}

Such a process can be executed in as little as 5 minutes (e.g. synchronously with Remesh), or over days or weeks (e.g. asynchronously and iteratively with Polis \cite{smallPolis2021}).
\end{example}

\hypertarget{system-design-properties-and-principles}{%
\section{Properties and Principles}\label{system-design-properties-and-principles}}

\paragraph{Key properties} The following properties must be satisfied for a system to be a collective response system:

\begin{enumerate}
\tightlist
\item
  \textbf{Participant Agency}: Participants themselves may suggest responses instead of being limited to a fixed set---this may be referred to as ``\emph{participatory perspectives''}.\footnote{This is what the WikiSurvey paper calls collaborativeness. \cite{salganikWikiSurveysOpen2015a}}\footnote{Responses may also be suggested from other means than the participants (e.g.~by facilitators, supporting generative models, etc.), but those are treated just like participant responses, and potentially flagged with information about their origin.} 
\item
  \textbf{Parallel Communication}: Every participant doesn't need to talk to or listen to every other in order to move the process forward. This may rely on methods such as \emph{bracketing} (i.e.~roughly analogous to breakout groups \cite{PSi}) and \emph{elicitation inference} (i.e.~methods for approximating how the participants would evaluate every response given evaluations of a subset) \cite{konya2022elicitation}.
\item
  \textbf{Representative Distillation}: There is some mechanism for approximating and/or aggregating the input of all of the participants to get evaluations representative of the whole. The resulting distilled outputs (which may include a ``map'' of output \cite{smallPolis2021}), are shared back to the participants.
\end{enumerate}

\label{key-principles}\paragraph{Key principles} In addition, the design of a collective response system must be designed to satisfy the following key principles in order to live up to the democratic ideals described earlier:

\begin{enumerate}
\tightlist
\item
  \textbf{Collective Oriented}: The system design prioritizes {supporting/understanding a group and its decision-making} \emph{over} supporting/understanding an individual and their decision-making.\footnote{For example, a normal recommendation system fails this test---it aims to give each user what they would want.}\footnote{Current deliberative polls may also arguably fail this test as they focus on individual choice changes instead of collective conclusions.}
\item
  \textbf{Prompt Oriented}: The system design prioritizes the creation of {outputs that best respond to the input prompt} \emph{over} other goals (e.g.~entertainment).\footnote{Under this principle, arguably, connection, trust-building, practice in deliberative communication, etc. might also be prioritized in some systems, but only in the service of better prompt outputs. This is potentially worth iterating on or providing exceptions for.}
\item
  \textbf{Deliberation Oriented}: The system design prioritizes affordances and incentives that support {participants understanding the potential choices---and each other's perspectives on them}\footnote{Including potentially the relative extent of the different perspectives.}---\emph{over} other goals (e.g.~engagement).\footnote{In other words, the purpose is to output a collective's response to a prompt in a way that \href{https://www.belfercenter.org/publication/holding-platforms-accountable-not-enough-we-need-compass-social-technologies}{supports mutual understanding, trust, and wise decision-making}; or alternatively ``combines \href{https://www.radicalxchange.org/media/blog/plurality-technology-for-collaborative-diversity-and-democracy/}{plurality} and \href{https://aviv.medium.com/building-wise-systems-combining-competence-alignment-and-robustness-a9ed872468d3}{wisdom}''.} (For example, this can involve supporting actions such as identifying common ground, surfacing insight, self-reflection on group experiences, and mapping of perspectives.)
\item
  \textbf{Fair Hearing Oriented}: The system design must ensure that all participants are given a fair hearing---{all responses are given the opportunity to be heard by the group such that they may end up being considered the best}.\footnote{While the ideal of the collective response system might involve fully equitable listening, depending on the system, responses with more traction may be seen more, and some responses may be shown for other reasons, e.g.~to better infer what has the most common ground. (Polis does this) It may also be helpful in some contexts to have a reputation score with some influence over the process.}
\end{enumerate}

Properties are binary and clearcut, while the principles leave more room for interpretation; there may be additional philosophical, empirical, and normative work to be done to refine these principles further.

\hypertarget{collective-dialogue}{%
\subsection{Collective Dialogue}\label{collective-dialogue}}

A \textbf{collective dialogue system} (CDS) is a collective response system with some form of iteration or feedback loop, such that participants repeatedly interact with the system to create new responses based on previous outputs (either to the same prompts or new prompts). The term \textbf{collective dialogue} can be used as shorthand for a collective dialogue process (the processes facilitated by a CDS).

Such a dialogue process enables further \emph{drilling down into details, motivations, and blockers} for the most promising solutions. They can also be used for any other sense-making or iterative decision-making task. 

\begin{example}
Consider the example earlier asking \emph{``What is the most important challenge in education facing our city?''}. If it was part of a collective dialogue, after the first answer, a follow-up question might be asked based on the responses that are deemed most representative, in a new collective response process with the same participants. \\
This might look like asking: \emph{``What do you think is the best solution for addressing {[}that challenge{]}?''}
\end{example}

\hypertarget{related-concepts}{%
\section{Existing Work \& Related Concepts}\label{related-concepts}}
There are many existing partial implementations of collective response systems and a few complete ones. Some of the most notable include \emph{\href{https://en.wikipedia.org/wiki/Wiki_survey}{WikiSurvey}'s} like \emph{\href{http://pol.is}{Polis}} and \emph{\href{https://www.allourideas.org/}{All Our Ideas}} \cite{smallPolis2021, salganikWikiSurveysOpen2015a}.\footnote{Unlike a WikiSurvey, collective response systems do not need to be `greedy', and some collective response systems intentionally limit what can be evaluated.} \emph{\href{https://www.remesh.ai/}{Remesh}} and similar tools focus on approximation components which enable greater scale \cite{2021ASRSGWilliamsConductsRemesh}. An alternative approach can be found in bracketing systems like \emph{\href{https://www.thepsiapp.com/}{PSiApp}} \cite{PSi}. Some processes created for \emph{citizens' assemblies} \cite{kimbrawhite2022FacilitatingDeliberationPractical} and \emph{deliberative polls} \cite{fishkin2019DemocracyWhenPeople} may also fulfill the criteria for a collective response system. Relatedly, \emph{sortition}-based selection of participants can be used to select people within a much larger group in situations where using collective response systems for the entire group is impractical.

Collective response systems can blur the lines between \emph{governance systems} and \emph{recommender systems}. For example, collective response systems may use methods such as \emph{matrix factorization} to support approximation \cite{bilichRemesh2019}.\footnote{For comparison, Reddit's comments and their ranking satisfies the three properties of a collective response system, though they do not satisfy most of the principles.} More recent approaches have also involved the use of large language models \cite{konya2022elicitation,bakker2022FinetuningLanguageModelsa}.

Other kinds of online machine learning systems are also likely to be related to collective response systems (particularly depending on where and how training data is sourced). The chaining of collective response systems together into complex decision-making networks can be somewhat analogous to both prompt chaining \cite{wu2022PromptChainerChainingLarge} and multi-body sortition \cite{bouriciusWhyHybridBicameralism2018}.\footnote{Such a \emph{multi-collective dialogue} might chain together several outputs. For example, group 1 identifies the most important problem to address, groups 2-4 then propose solutions to that problem, group 5 identifies issues with the proposed solutions and key tradeoffs, group 6 integrates the most important results of the earlier groups into a single solution, and group 7 does a final `quality control' check on the proposed solution. Any of these collective dialogues may also be replaced with other kinds of processes as is appropriate to the context, from citizens' assemblies to foundation model inferences.} There may also be relevant similarities between prompt `engineering' for generative AI and prompt design for generative CI.

More generally, collective response systems are often meant to address multi-principal-agent problems \cite{konya2022elicitation}, so prior work related to such problems is likely to be relevant. Finally disciplines including social choice theory have much to say on potential definitions of `best'---though the insight of other disciplines can also be incorporated. This is clearly not an exhaustive list, but simply meant to illustrate part of the breadth of related work.

\hypertarget{conclusion}{%
\section{Conclusion}\label{conclusion}}

The underlying ideas and ideals of collective response systems are not new---this paper is primarily meant to articulate them succinctly, with the goal of creating a useful shared language.

Collective response systems support understanding of a group's values, needs, and desires by combining participant-led responses and evaluations with a reversal of attention economy incentives---which then helps create decisive mandates. Such systems can be used to identify and reward responses that can bridge divides---the opposite of most modern politics, media, and social media. Anecdotally, collective response systems can also be incredibly efficient at surfacing new insights from those closest to the issues, and people feel heard and respected when they see their responses incorporated into the process.

\subsection{Limitations}
Collective response systems are not a panacea. 

An isolated collective response process will only elicit gut opinions, not considered judgments. This is somewhat addressed in more elaborate (and expensive) processes such as citizens assemblies and deliberative polls through access to educational materials, experts, stakeholders, compensated time, and careful neutral moderation \cite{fishkin2019DemocracyWhenPeople, kimbrawhite2022FacilitatingDeliberationPractical, oecdInnovativeCitizenParticipation2020a}. Such resourcing and deliberative depth is crucial for decisions of significant importance. 

Some (but not all) of the downsides of the collective response system can be overcome with iteration, i.e. collective dialogues. Such iteration can enable deeper and more deliberative explorations. Iterated or chained collective dialogues can even be used to determine e.g. what questions are most important to ask experts, stakeholders, or the entire group.\footnote{They may also potentially be used to identify tradeoffs and `cruxes' that can then be resolved.}

However, even when using a well-designed collective dialogue system, careful process design is crucial. Particular collective response systems implementations may also have significant tradeoffs, e.g. in their capacity to be chained together into dialogues, to incorporate new information from participants mid-process, to be synchronous vs. asynchronous, to involve text vs. audio or video, etc. All of these factors must be taken into account.

Many digitally mediated collective response systems do not, by default, enable normal human relationship formation (similar to ordinary, non-generative, voting systems). However, they do (surprisingly) appear to enable a valuable form of connection and trust despite that, likely by facilitating a form of \emph{collective introspection} and helping identify areas of common ground through representative distillation. %

Where possible, collective response processes would ideally be used as a complement to deeper deliberative processes instead of a replacement. For example, they might be used by a citizens' assembly to get open-ended public input from the public in order to support its decision-making. Due to their lower overheard, collective response processes may also be run on a regular basis, far more often than assemblies, for less important decisions or for specific components of a governance process where deep personal deliberation is less crucial.

\subsection{Opportunities}
While not appropriate for all collective decisions, collective response systems may help for embed new forms of sense-making, governance, democracy, and agency at every level of society:

\begin{itemize}
\tightlist
\item
  \textbf{Governance \& Conflict}: They can be deployed by representatives, civil servants, citizen assemblies, and conflict mediators to meaningfully include the broader (busy) public.
\item
  \textbf{`Corporate democracy'}: They support a new form of corporate governance, going beyond just shareholders and elite stakeholders to users, employees, and the impacted public.
\item
  \textbf{Media \& Understanding}: They can support a new form of collective introspection---helping a public `know itself', identify common ground, and thus better navigate internal and external challenges.
\end{itemize}

There is a significant opportunity for research to create evaluation protocols and metrics for such systems, understand the impacts of different design decisions, and thus develop increasingly better systems (and chains of systems) for different purposes and contexts. 

Just as advances in machine learning have lead to new ways of interacting with computers by enabling `generative AI', advances in collective response systems can lead to new ways of interacting with groups by supporting `generative collective intelligence' (CI).

\section*{Acknowledgements}
Thanks to Anna Waldman-Brown, Andrew Konya, Colin Megill, Divya Siddarth, Ted Suzman, Bruce Schneier, Luke Thorburn, Antoine Vergne, Glen Weyl, Kimbra White, Cathy Wu, Jessica Yu, and others for the conversations and feedback that resulted in this paper.

\bibliographystyle{unsrt}  
\bibliography{references}  %

\end{document}